\documentclass[a4paper,twocolumn,11pt,accepted=2026-03-24]{quantumarticle}
\pdfoutput=1
\usepackage[utf8]{inputenc}
\usepackage[english]{babel}
\usepackage[T1]{fontenc}
\usepackage{amsmath}
\usepackage{hyperref}
\usepackage{amsfonts}
\usepackage{amssymb}
\usepackage{amsmath}
\usepackage{braket}
\usepackage{caption}
\usepackage{subcaption}
\usepackage{graphicx}

\usepackage{tikz}
\usepackage{lipsum}

\newtheorem{theorem}{Theorem}

\begin{document}

\title{A Remarkable Application of Zassenhaus Formula to Strongly Correlated Electron Systems}

\author{Louis Jourdan}
\affiliation{Universit\'e C\^ote d'Azur, CNRS, LJAD, UMR 7351, 28 avenue Valrose, 06108 Nice, France}
\orcid{0009-0000-0587-5912}
\author{Patrick Cassam-Chenaï}
\email{cassam@unice.fr}
\homepage{https://math.univ-cotedazur.fr/~cassam/}
\orcid{0000-0002-9437-7794}
\affiliation{Universit\'e C\^ote d'Azur, CNRS, LJAD, UMR 7351, 28 avenue Valrose, 06108 Nice, France}

\maketitle

\onecolumn
\begin{abstract}
  We show that the Zassenhaus decomposition for the exponential of the sum of two non-commuting operators, simplifies drastically when these operators satisfy a simple condition, called the no-mixed adjoint property. An important application to a Unitary Coupled Cluster method for strongly correlated electron systems is presented. This ansatz requires no Trotterization and is exact on a quantum computer with a finite number of Givens gate equals to the number of free parameters. The formulas obtained in this study also shed light on why and when optimization after Trotterization gives exact solutions in disentangled forms of unitary coupled cluster.
\end{abstract}

\section{Introduction}\label{sec1}

Quantum algorithms based on the Zassenhaus product formula is an active field of research~\cite{Casas2025}.
The Zassenhaus formula is a kind of dual formula of the celebrated Baker-Campbell-Hausdorff (BCH) one~\cite{Magnus1954}. It decomposes the exponential of the sum of two non-commuting operators, $\hat{X}$ and $\hat{Y}$, into a product of exponentials of homogeneous Lie polynomials of increasing degrees:
\begin{equation}
    \exp(\hat{X}+\hat{Y})=\exp(\hat{X})\exp(\hat{Y})\prod^{+\infty}_{n=2}\exp(\hat{C}_{n}(\hat{X},\hat{Y})),
    \label{zassenhaus}
\end{equation}

where $\hat{C}_{n}(\hat{X},\hat{Y})$ is a homogeneous Lie polynomial of degree $n$ in $\hat{X}$ and $\hat{Y}$. This decomposition is unique and recursive formulas for the $\hat{C}_{n}(\hat{X},\hat{Y})$ polynomials have been given by Wilcox~\cite{Wilcox1967}, the latter indicating also a simpler method followed later by Suzuki~\cite{Suzuki1977}. Since then, various approaches have been developed in view of computer implementations, see~\cite{Scholz2006,Casas2012,Wang2019} to quote a few,  a more comprehensive bibliography can be found in~\cite{Casas2012}. Note that the Zassenhaus formula can also be derived from the Fer-Wilcox expansion~\cite{Fer1958,Wilcox1967,Arnal2021}.\\

This article is concerned with applications of the Zassenhaus formula to traditional quantum mechanics, although Zassenhaus recursions also appear in quantum field theory~\cite{Ebrahimi-Fard2008}.
Decomposing the exponential of non-commuting operators is of paramount importance in quantum physics and quantum computing.  It is usually done by iterating a truncated version of the Zassenhaus formula at order, say $m$, after having rewritten the sum of operators as $k\cdot\left(\frac{\hat{X}}{k}+\frac{\hat{Y}}{k}\right)$:
\begin{equation}
    \exp(\hat{X}+\hat{Y})=\lim_{k\rightarrow\infty} \left[\exp\left(\frac{\hat{X}}{k}\right)\exp\left(\frac{\hat{Y}}{k}\right)\prod^{m}_{n=2}\exp\left(\hat{C}_{n}(\frac{\hat{X}}{k},\frac{\hat{Y}}{k})\right)\right]^k,
    \label{truncated_zassenhaus}
\end{equation}
The most widespread such strategy consists in just applying Trotter's formula~\cite{Trotter1959}, which corresponds to the shortest non-trivial truncation of the Zassenhaus formula:
\begin{equation}
    \exp(\hat{X}+\hat{Y})=\lim_{k\rightarrow\infty} \left[\exp\left(\frac{\hat{X}}{k}\right)\exp\left(\frac{\hat{Y}}{k}\right)\right]^k.
    \label{trotter}
\end{equation}
In practice, the number, $k$, of Trotter steps is finite,
\begin{equation}
    \exp(\hat{X}+\hat{Y})\approx \left[\exp\left(\frac{\hat{X}}{k}\right)\exp\left(\frac{\hat{Y}}{k}\right)\right]^k,
    \label{truncated_trotter}
\end{equation}
 and the word "Trotterization" has now become part of the common language of quantum computing, to designate this process.\\

Trotterization is an approximation. Even if the  Zassenhaus formula is truncated at a higher order, the stepsize is always finite in real life computation, and actually often kept as small as $k=1$ to limit the number of quantum gates on Noisy Intermediate-Scale Quantum (NISQ) computers~\cite{Preskill2018}. This is an extra source of loss of accuracy. So, it is highly desirable to find short, exact, alternative decompositions of $\exp(\hat{X}+\hat{Y})$.\\

The purpose of the present article is to provide such a decomposition in a particular, yet potentially very useful, case of operators representing mono- and di-excitation of electronic states in quantum physics and chemistry. Note that there are already known cases of such closed-form decomposition in Lie algebra theory~\cite{Kurlin2007}, see also Ref.~\cite{jayakumar2025} for a similar endeavor focusing on the BCH formula.\\ 

The article is organized as follows: In the next section, we demonstrate the simplifications which occur in the Zassenhaus formula, when the iterated commutators of abstract $\hat{X}$ and $\hat{Y}$ operators satisfy a specific property, called the "no-mixed adjoint" property. Then, we show that the  mono- and di-excitation operators relevant to a promising ansatz for electronic wave functions, satisfy this property, so that the simplified Zassenhaus decomposition can be applied to it, and translations into quantum circuits are proposed. We conclude with the consequences of this finding for the implementation of the ansatz on quantum computers.

\section{Zassenhaus formula for no-mixed adjoint operators}

We define the iterated adjoint of an operator, acting typically on the Hilbert space of a quantum system, $\hat{X}$ with respect to another such operator $\hat{Y}$ by recursion:
\begin{align}
ad^{0}_{\hat{X}}\hat{Y}&=&\hat{Y}\\
\forall j\in\mathbb{N}^{*},\quad ad^{j}_{\hat{X}}\hat{Y}&=&[\hat{X},ad^{j-1}_{\hat{X}}\hat{Y}],
\label{iterated_adjoint}
\end{align} 
where the bracket $[\cdot , \cdot]$, denotes the commutator between two operators.
The iterated adjoint $ad^{j}_{\hat{X}}$ can be regarded as a mapping which provides a useful notation for nested commutators with $\hat{X}$. In particular, $ad^{1}_{\hat{X}}\hat{Y}\equiv ad_{\hat{X}}\hat{Y} =[\hat{X},\hat{Y}]$.
We use the iterated adjoint in the next subsection, to introduce the, so-called, "no-mixed adjoint property".

\subsection{No-mixed adjoint property}
We say that the couple of operators $(\hat{X},\hat{Y})$ satisfies the no-mixed adjoint property, if:
\begin{equation}
   \forall i\in\mathbb{N}\quad    ad_{\hat{Y}}ad^{i}_{\hat{X}}\hat{Y}=0
   \label{no-mixed}
\end{equation} (which implies that all chains of adjoints $ ad^{j_{m}}_{\hat{Y}}ad^{i_{m}}_{\hat{X}}\ldots ad^{j_{1}}_{\hat{Y}}ad^{i_{1}}_{\hat{X}}\hat{Y}$ are  zero, unless\\ $j_1=\cdots=j_m=0$, that is to say, the chain is non-mixed and can be written more simply as $ad^{i}_{\hat{X}}\hat{Y}$ with $i=i_m+\cdots+i_1$).
With this definition, one can prove the following theorem.\\

\begin{theorem}
\label{theorem}
    Let $(\hat{X},\hat{Y})$ satisfying the no-mixed adjoint property, then the homogeneous Lie polynomials $\hat{C}_{n}(\hat{X},\hat{Y})$ of the Zassenhaus formula for these operators,  Eq.(\ref{zassenhaus}), are given by:
    \begin{equation}
        \forall n\geqslant2 \quad \hat{C}_{n}(\hat{X},\hat{Y})=\frac{(-1)^{n-1}}{n!}ad^{n-1}_{\hat{X}}\hat{Y} .
    \label{simpleC}
    \end{equation}
    
\end{theorem}

The detailed proof of this theorem can be found in Appendix A.

\subsection{Zassenhaus decomposition:}
This last expression can be extended to $n=1$, with $\hat{C}_{1}(\hat{X},\hat{Y})=\hat{Y}$, so that the Zassenhaus decomposition writes:

\begin{align}
    \exp(\hat{X}+\hat{Y})&=&\exp(\hat{X})\prod^{+\infty}_{n=1}\exp(\hat{C}_{n}(\hat{X},\hat{Y}))\nonumber\\
    &=&\exp(\hat{X})\prod^{+\infty}_{n=1}\exp\left(\frac{(-1)^{n-1}}{n!}ad^{n-1}_{\hat{X}}\hat{Y}\right)\nonumber\\
    &=&\exp(\hat{X})\exp\left(\sum\limits^{+\infty}_{n=1}\frac{(-1)^{n-1}}{n!}ad^{n-1}_{\hat{X}}\hat{Y}\right),
    \label{zassenhaus-simple}   
\end{align}

where  we have made use of the corollary, Eq.(\ref{conseqlemme1}), of the Appendix A, for the last equality.\\
The series in the second exponential appears in the formula of the derivative of the exponential map, in a Lie algebra context. It can be denoted formally as $\frac{1-\exp(-ad_{\hat{X}}\hat{Y})}{ad_{\hat{X}}\hat{Y}}$, and replaced by an integral,
\begin{align}
\sum\limits^{+\infty}_{n=1}\frac{(-1)^{n-1}}{n!}ad^{n-1}_{\hat{X}}\hat{Y}&=&\int_0^1\ \exp(ad_{-t\hat{X}})\hat{Y}dt\ -1  \nonumber\\
&=&\int_0^1\ \exp(-t\hat{X})\cdot\hat{Y}\cdot \exp(t\hat{X}) dt\ -1.
    \label{duhamel}   
\end{align}
Its inverse also appears in the differential equation leading to Magnus expansion~\cite{Magnus1954,Bauer2013}.\\ A last side-remark is that, if an operator $\hat{Y}'$ can be cast in the form  $\hat{Y}'=\frac{1-\exp(-ad_{\hat{X}}\hat{Y})}{ad_{\hat{X}}\hat{Y}}$ with the no-mixed adjoint property satisfied between $\hat{X}$ and $\hat{Y}$, then the Baker-Campbell-Hausdorff expansion for $\hat{X}$ and $\hat{Y}'$ will simplify, as $\exp{\hat{X}}\exp{\hat{Y}'}=\exp{(\hat{X}+\hat{Y})}$.

\section{Application to a Unitary Coupled Cluster method}
 
\subsection{Unitary 2D-Block Frozen Pair Coupled Clusters}
The coupled cluster (CC) method has its origin in nuclear physics and has now spread in many branches of physics and chemistry~\cite{Bartlett2007,Laestadius2019}. It is sometimes qualified as the "gold-standard" of quantum chemistry. The CC ansatz consists in optimizing parametrized wave functions of the form $\phi=\exp(\hat{T})\phi_0$, where $\phi_0$ is a fixed wave function called the "reference" wave function, and  $\hat{T}$ is a parametrized operator called  the "cluster" operator. When the latter is split into two commuting parts $\hat{T}=\hat{T}_{int}+\hat{T}_{ext}$, and the reference wave function is a so-called "Slater determinant",  then one obtains a tailored coupled cluster (tCC) ansatz~\cite{Leszczyk2022,Tecmer2022}. Besides, if $\hat{T}$ is anti-Hermitian, so that $\exp(\hat{T})$ is unitary, when the parameters of $\hat{T}$ are optimized variationally, one talks about the unitary coupled cluster (UCC) method. Recently, UCC has attracted a renewed interest in quantum computing~\cite{Peruzzo14,Yung14,McClean2016,Romero2019}  and has been developed into many different forms~\cite{Grimsley2019,Lee2019,Sokolov2020,Huggins2020,Ryabinkin2020,Xie2022}, to quote a few.\\

A promising ansatz for strongly correlated electron systems, which combines tCC and UCC is the unitary frozen pair coupled cluster (UfpCC) ansatz, where 
$\hat{T}_{int}$ is of the form, in second quantization formalism,
\begin{equation}
\hat{T}_{int}   =  \sum^{n_{occ}}_{p=1}\sum_{q=n_{occ} +1}^{n_{orb}}\lambda_{pq}(\hat{S}^{+}_{q}\hat{S}^{-}_{p}-\hat{S}^{+}_{p}\hat{S}^{-}_{q})
\label{pair-op}
\end{equation}
where the $\lambda_{pq}$'s are real numbers, $\hat{S}^{+}_{i}$ is an operator "creating" an electron pair, $\hat{S}^{-}_{i}=(\hat{S}^{+}_{i})^\dagger$ its Hermitian conjugate "annihilating" an electron pair, $n_{occ}$ is the number of occupied electron pairs in the reference, $\ket{\phi_0}=\prod\limits^{n_{occ}}_{p=1}\hat{S}^{+}_{p}\ket{0}$, and $n_{orb}$ is the total number of "orbitals" i.e. of spinless one-electron states. Different choices are possible for $\hat{T}_{ext}$. The simplest one is to limit $\hat{T}_{ext}$ to broken-pair double excitations $\hat{T}_{ext}=\hat{T'}_{2}$. However, one might wish to include also single-excitations $\hat{T}_{ext}=\hat{T}_{1} + \hat{T'}_{2}$  \cite{Tecmer2022}. Then, in Unitary frozen pair Single and Double Coupled Cluster (UfpSDCC),  the Zassenhaus formula can prove useful, because
\begin{equation}
\hat{T}_{1}   =  \sum^{n_{occ}}_{p=1}\sum_{q=n_{occ} +1}^{n_{orb}}\mu_{pq}(\hat{S}^{+}_{pq}\hat{S}^{-}_{p}-\hat{S}^{+}_{p}\hat{S}^{-}_{pq})
\label{T1}
\end{equation}
and 
\begin{equation}
\hat{T'}_{2}   =  \sum^{n_{occ}}_{1\leq p_1<p_2}\ \sum_{n_{occ} +1\leq q_1\neq q_2}^{n_{orb}}\mu_{p_1p_2q_1q_2}(\hat{S}^{+}_{p_1q_1}\hat{S}^{+}_{p_2q_2}\hat{S}^{-}_{p_2}\hat{S}^{-}_{p_1}-\hat{S}^{+}_{p_1}\hat{S}^{+}_{p_2}\hat{S}^{-}_{p_2q_2}\hat{S}^{-}_{p_1q_1})
\label{T2}
\end{equation}
do not commute. In Eqs.(\ref{T1}) and (\ref{T2}), the $\mu$'s are real parameters. The non-commutation arises from the pair operators. The latter are expressed in terms of one-electron creation and annihilation operators, 
\begin{equation} 
\hat{S}^{+}_{ij}=\frac{1}{\sqrt{2(1+\delta_{ij})}}(\hat{a}^{\dagger}_{i+\frac{1}{2}}\hat{a}^{\dagger}_{j-\frac{1}{2}}+\hat{a}^{\dagger}_{j+\frac{1}{2}}\hat{a}^{\dagger}_{i-\frac{1}{2}})=(\hat{S}^{-}_{ij})^{\dagger},
\end{equation}
\begin{equation} 
\hat{S}^{-}_{ij}=\frac{1}{\sqrt{2(1+\delta_{ij})}}(\hat{a}_{j-\frac{1}{2}}\hat{a}_{i+\frac{1}{2}}+\hat{a}_{i-\frac{1}{2}}\hat{a}_{j+\frac{1}{2}})=(\hat{S}^{+}_{ij})^{\dagger},
\end{equation}
where $\hat{a}_{i\sigma}^\dagger$ (resp. $\hat{a}_{i\sigma}$) creates (resp. annihilates) an electron in the spin-orbital of orbital part $i$ and spin $\sigma$, these operators obeying the canonical anticommutation relations, $[\hat{a}_{i\sigma},\hat{a}_{j\tau}]_+=[\hat{a}_{i\sigma}^\dagger,\hat{a}_{j\tau}^\dagger]_+=0$, $[\hat{a}_{i\sigma},\hat{a}_{j\tau}^\dagger]_+=\delta_{ij}\delta_{\sigma\tau}$. The notation can be simplified for identical indices: $\hat{S}^{-}_i\equiv\hat{S}^{-}_{ii}$ and $\hat{S}^{+}_{i}\equiv\hat{S}^{+}_{ii}$.\\

Based on our previous findings \cite{Cassam_hal-05210807} we advocate a simplified "$2D$-Block" version of the UfpSDCC ansatz, where each orbital $p_i$ occupied by an electron pair, is broken by means of an excitation towards only one, specifically associated, virtual orbital, $q_i$, reminiscent of the generalized valence bond perfect-pairing (GVB-PP) method~\cite{Goddard1967,Goddard1973}. So, it can also be regarded as an UCC analogue of a Block-Correlated Coupled-Cluster (BCCC) method~\cite{Wang2020} but with a different technique to partition the orbitals into blocks.
The couple $(p_i,q_i)$ forms, what we call the $i^{th}$ $2D$-block. Note that, this orbital partitioning into blocks is only relevant when it has been optimized in a specific way~\cite{Cassam_hal-05210807}. Then, the $\hat{T}_{1}$ cluster operator becomes,
\begin{equation}
\hat{Y}  =  \sum^{n_{occ}}_{i=1}\mu_{i}(\hat{S}^{+}_{p_iq_i}\hat{S}^{-}_{p_i}-\hat{S}^{+}_{p_i}\hat{S}^{-}_{p_iq_i}),
\end{equation}
and the $\hat{T'}_{2}$
\begin{equation}
\hat{X}   =  \sum^{n_{occ}}_{1\leq i<j}\mu_{ij}(\hat{S}^{+}_{p_iq_i}\hat{S}^{+}_{p_jq_j}\hat{S}^{-}_{p_j}\hat{S}^{-}_{p_i}-\hat{S}^{+}_{p_i}\hat{S}^{+}_{p_j}\hat{S}^{-}_{p_jq_j}\hat{S}^{-}_{p_iq_i}).
\end{equation}
Although  more simple, they still do not commute. More precisely, setting 
\begin{align}
 \hat{A}_{ij}&=&\hat{S}^{+}_{p_iq_i}\hat{S}^{+}_{p_jq_j}\hat{S}^{-}_{p_j}\hat{S}^{-}_{p_i}-\hat{S}^{+}_{p_i}\hat{S}^{+}_{p_j}\hat{S}^{-}_{p_jq_j}\hat{S}^{-}_{p_iq_i}   \\
 \hat{B}_i&=&\hat{S}^{+}_{p_iq_i}\hat{S}^{-}_{p_i}-\hat{S}^{+}_{p_i}\hat{S}^{-}_{p_iq_i}
\end{align} 
we have the following commutation relations,
\begin{align}
 \forall i,j,k,l \qquad [\hat{A}_{ij},\hat{A}_{kl}] &=& [\hat{B}_{i},\hat{B}_{k}] = 0 ,
 \label{A-or-B-comm}
\end{align} 
\begin{align}
 \forall i,j,k \qquad 
[\hat{A}_{ij},\hat{B}_{k}] &=& \delta_{ik} \hat{B}_{j} + \delta_{jk} \hat{B}_{i}.
\label{A-and-B-comm}
\end{align} 

Then, we easily obtain that,  $\hat{Y}=  \sum\limits^{n_{occ}}_{i=1}\mu_{i}\hat{B}_{i}$, and $\hat{X}=  \sum\limits^{n_{occ}}_{1\leq i<j}\mu_{ij}\hat{A}_{ij}$  fulfill the no-mixed adjoint property, Eq.(\ref{no-mixed}), 
since, according to Eqs.(\ref{A-or-B-comm}) and (\ref{A-and-B-comm}), $ad^{i}_{\hat{X}}\hat{Y}$ can only depend on $\hat{B}_{k}$ operators, which all commute with $\hat{Y}$. So,  the simplified formula, Eq.(\ref{simpleC}), for the Zassenhaus coefficients does apply.

\subsection{Case of two electron pairs}

The simplest non-trivial case, $n_{occ}=2$, corresponds to a two electron-pair system, such as the lithium hydride, $LiH$, molecule. 
Then, $\hat{X}=\mu_{12}\hat{A}_{12}$ and $\hat{Y}=\mu_{1}\hat{B}_{1}+\mu_{2}\hat{B}_{2}$, the $\mu$'s parameters being non-zero. In fact, one could add more $\hat{B}_{k}$, $k>2$, operators in $\hat{Y}$, it would not add any difficulty, as these additional operators would commute with everything else.\\

We have to calculate the iterated adjoints,  $ad^{n}_{\hat{X}}\hat{Y}$, to be able to use Eq.(\ref{simpleC}). Two cases are to be distinguished depending on the parity of $n$. For $n=2k$ even, iterating Eq.(\ref{A-and-B-comm}), taking into account additional $\mu$-coefficients, one has:
\begin{align}
ad^{2k}_{\hat{X}}\hat{Y}&=&(\mu_{12})^{2k}(\mu_{1}\hat{B}_{1}+\mu_{2}\hat{B}_{2}),
\label{even}
\end{align}
while for  $n=2k+1$ odd,
\begin{align}
ad^{2k+1}_{\hat{X}}\hat{Y}&=&(\mu_{12})^{2k+1}(\mu_{1}\hat{B}_{2}+\mu_{2}\hat{B}_{1}).
\label{odd}
\end{align}
This leads to the following Zassenhaus coefficients, $\forall k\geq 1$,
\begin{equation}
    \hat{C}_{2k}(\hat{X},\hat{Y})=\frac{-1}{(2k)!}(\mu_{12})^{2k-1}(\mu_{1}\hat{B}_{2}+\mu_{2}\hat{B}_{1}),
    \label{simpleC-even}
\end{equation}
\begin{equation}
    \hat{C}_{2k+1}(\hat{X},\hat{Y})=\frac{1}{(2k+1)!}(\mu_{12})^{2k}(\mu_{1}\hat{B}_{1}+\mu_{2}\hat{B}_{2}).
    \label{simpleC-odd}
\end{equation}
Their sum splits according to parity, and then can be factored into a $\hat{B}_{1}$ and a $\hat{B}_{2}$ term, 
\begin{align}
    \sum^{+\infty}_{n=1}\hat{C}_{n}(\hat{X},\hat{Y})&=&\sum^{+\infty}_{k\geq 0}\frac{1}{(2k+1)!}(\mu_{12})^{2k}(\mu_{1}\hat{B}_{1}+\mu_{2}\hat{B}_{2}) + \sum^{+\infty}_{k> 0}\frac{-1}{(2k)!}(\mu_{12})^{2k-1}(\mu_{1}\hat{B}_{2}+\mu_{2}\hat{B}_{1})\nonumber\\
    &=&\frac{\sinh(\mu_{12})}{\mu_{12}}(\mu_{1}\hat{B}_{1}+\mu_{2}\hat{B}_{2}) - \frac{\cosh(\mu_{12})-1}{\mu_{12}}(\mu_{1}\hat{B}_{2}+\mu_{2}\hat{B}_{1})\nonumber\\
&=&\frac{\sinh(\mu_{12})\mu_{1}-(\cosh(\mu_{12})-1)\mu_{2}}{\mu_{12}}\hat{B}_{1}+\frac{\sinh(\mu_{12})\mu_{2}-(\cosh(\mu_{12})-1)\mu_{1}}{\mu_{12}}\hat{B}_{2}\nonumber\\
\end{align}
So, the Zassenhaus decomposition gives:
\begin{align}
    \exp(\hat{X}+\hat{Y})=\exp(\mu_{12}\hat{A}_{12})\exp\left(\frac{\sinh(\mu_{12})\mu_{1}-(\cosh(\mu_{12})-1)\mu_{2}}{\mu_{12}}\hat{B}_{1}+\right.\nonumber\\\left.\frac{\sinh(\mu_{12})\mu_{2}-(\cosh(\mu_{12})-1)\mu_{1}}{\mu_{12}}\hat{B}_{2}\right).
    \label{zassenhaus-2pairs}
\end{align}
The $\hat{B}_{i}$'s commuting, this can be re-expressed as a product of exponentials. If more $\hat{B}_{k}$, $k>2$, are present, one can write more generally:
\begin{align}
    \exp(\hat{X}+\hat{Y})=\exp(\mu_{12}\hat{A}_{12})\exp\left(\frac{\sinh(\mu_{12})\mu_{1}-(\cosh(\mu_{12})-1)\mu_{2}}{\mu_{12}}\hat{B}_{1}\right)\nonumber\\\exp\left(\frac{\sinh(\mu_{12})\mu_{2}-(\cosh(\mu_{12})-1)\mu_{1}}{\mu_{12}}\hat{B}_{2}\right)\prod_{k>2}\exp(\mu_{k}\hat{B}_{k}).
    \label{zassenhaus-2pairs-ext}
\end{align}
This is a property of paramount importance, especially in view of quantum computing applications, as it saves an \textit{a priori} infinite number of quantum gates to achieve an exact unitary transformation. Only the generating operators $A_{ij}$ and $B_k$ need to be implemented as quantum circuits. This extends to the general case.\\

In real life applications, the cluster operators are used to correct an already reasonable reference state. So, the $\mu$'s parameters are supposed to be small. An important remark is that, as $\mu_{12}\rightarrow 0$, $\left(\frac{\sinh(\mu_{12})\mu_{1}-(\cosh(\mu_{12})-1)\mu_{2}}{\mu_{12}}\hat{B}_{1}\right)\rightarrow\mu_{1}\hat{B}_{1}$ and $\left(\frac{\sinh(\mu_{12})\mu_{2}-(\cosh(\mu_{12})-1)\mu_{1}}{\mu_{12}}\hat{B}_{2}\right)\rightarrow\mu_{2}\hat{B}_{2}$, that is to say, although the commutator $ad^{1}_{\hat{X}}\hat{Y}=\mu_{12}(\mu_{1}\hat{B}_{2}+\mu_{2}\hat{B}_{1})$ is of order $1$ in $\mu_{12}$, it is as if $\hat{X}$ and $\hat{Y}$ were commuting to first order in Eq.(\ref{zassenhaus-2pairs-ext}). This also implies that, if we set $\alpha=\mu_{12},\beta=\frac{\sinh(\mu_{12})\mu_{1}-(\cosh(\mu_{12})-1)\mu_{2}}{\mu_{12}},\gamma=\frac{\sinh(\mu_{12})\mu_{2}-(\cosh(\mu_{12})-1)\mu_{1}}{\mu_{12}}$, then the Jacobian of the transformation is non-zero, so that, in practice, one can equivalently and more simply optimize $\alpha,\beta,\gamma$, instead of $\mu_{12},\mu_{1},\mu_{2}$. This also extends to the general case.\\

\begin{figure}[htpb]
    \centering 

\begin{subfigure}{0.4\textwidth}
\includegraphics[scale=0.7]{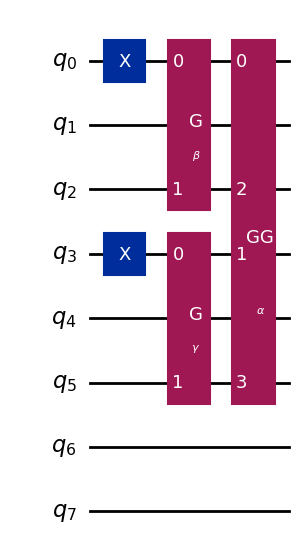}
    \caption{4-qubit Givens gate circuit}
    \label{circ1}
\end{subfigure}
\hfill
\begin{subfigure}{0.4\textwidth}
\includegraphics[scale=0.7]{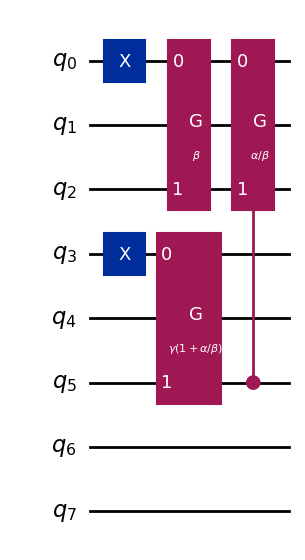}
    \caption{Conditional 2-qubit Givens gate circuit}
    \label{circ2}
\end{subfigure}
\caption{Two possible abstract circuits for a cluster operator, made of two mono- and one di-excitation for the LiH molecule in a minimal basis set. The $8$ qubits represents the following electron pairs in $6$ APSG-optimized orbitals, as advocated in \cite{Cassam_hal-05210807}, for the 2D-block geminal method: $\hat{S}^{+}_{1}\ket{0}$, $\hat{S}^{+}_{2}\ket{0}$, $\hat{S}^{+}_{12}\ket{0}$, $\hat{S}^{+}_{3}\ket{0}$, $\hat{S}^{+}_{4}\ket{0}$, $\hat{S}^{+}_{34}\ket{0}$, $\hat{S}^{+}_{5}\ket{0}$, $\hat{S}^{+}_{6}\ket{0}$. The last two qubits correspond to the lithium doubly occupied $p_x$ and $p_y$ orbitals, whose populations are optimized only in the reference wave function through a closed-shell pair unitary operator (not represented). They are not affected by the broken-pair cluster operator within the 2D-block ansatz, and the associated open-shell qubit $\hat{S}^{+}_{56}\ket{0}$ is not necessary. The left-hand side  circuit makes use of two and four-qubit Givens gates, while the  right-hand side one uses a conditional two-qubit Givens gate in place of the four-qubit Givens gate. Note that, this implies algebraic dependencies among the angle parameters, supposed small. The circuit drawings are exported from the IBM Quantum Qiskit packages and according to their convention, in practice, the sign of the gate angles should be reversed with respect to those appearing in the main text.}
\label{abstract-circuits}
\end{figure}

As shown by Arrazola et al.~\cite{Arrazola2022}, Givens rotations form a universal set of quantum circuits for quantum chemistry. However, the question arises whether to use controlled or uncontrolled Givens excitation gates.
Two possible implementations of the parametrized unitary operators of Eq.(\ref{zassenhaus-2pairs-ext}) in terms of quantum circuits are proposed in Fig. \ref{abstract-circuits}, for the elementary case $N=2$, with and without a controlled single-excitation gate. Their transpiled versions on an IBM Heron quantum computer are compared in Fig. \ref{real-circuits}. Although the use of a conditional two-qubit Givens gate involves only an abstract three-qubit gate, it induces algebraic dependencies among the parameters and the transpilation results in a deeper circuit than the one obtained by using a four-qubit Givens gate. Note that we do not use the usual spin-orbital encoding into qubits, instead each qubit encodes the occupation of an elementary electron pair. Our encoding generalizes that of Refs.~\cite{Khamoshi20,Elfving21} to include open-shell pairs and will be described in a forthcoming publication.\\

As we shall see in the next section, the structure of the ansatz in the general case is a simple generalization of the $N=2$ case, so the elementary circuits of Fig.\ref{abstract-circuits} can be used as building blocks for the preparation of a general ansatz.

\subsection{General Case}
Now, consider a system of $N=n_{occ}$ electron pairs,
\begin{equation}
\hat{X}=  \sum\limits_{1\leq i<j\leq N}\mu_{ij}\hat{A}_{ij},
\label{X}
\end{equation}
\begin{equation}
\hat{Y}=  \sum\limits^{N}_{i=1}\mu_{i}\hat{B}_{i},
\label{Y}
\end{equation}
the $\mu$'s parameters being non-zero, otherwise the corresponding operator could just be eliminated. 
We denote by $\mathcal{D}_{N,2}$ the set of $N(N-1)$ directed edges of $N$ vertices $\left\lbrace 1,\ldots,N\right\rbrace $.  So, if $i\rightarrow j\in\mathcal{D}_{N,2}$, we set $\nu_{i\rightarrow j}=\mu_{ij}$ if $i<j$, else if   $j<i$, $\nu_{i\rightarrow j}=\mu_{ji}$.
This notation is more convenient to express the general iterated adjoint,
\begin{align}
ad^{k}_{\hat{X}}\hat{Y}&=&\sum_{\stackrel{1\leq j_0,i_{k+1}\leq N,\ i_1\rightarrow j_1,\cdots, i_k\rightarrow j_k\in\mathcal{D}_{N,2}}{j_0=i_1, \ldots, j_k=i_{k+1}}}\mu_{j_0}\nu_{i_1\rightarrow j_1}\cdots\nu_{i_k\rightarrow j_k}\hat{B}_{i_{k+1}}.
\label{ad-gen}
\end{align}
The sum extends over all possible chains of indices such that two successive $\nu$-coefficients have at least one common index, starting with $j_0$, the single index of a $\mu$-coefficient, and ending with  $i_{k+1}$, the  index of a $\hat{B}$-operator. The Zassenhaus coefficients are expressed as,
\begin{equation}
    \hat{C}_{k+1}(\hat{X},\hat{Y})=\frac{(-1)^k}{(k+1)!}\sum_{\stackrel{1\leq j_0,i_{k+1}\leq N,\ i_1\rightarrow j_1,\cdots, i_k\rightarrow j_k\in\mathcal{D}_{N,2}}{j_0=i_1, \ldots, j_k=i_{k+1}}}\mu_{j_0}\nu_{i_1\rightarrow j_1}\cdots\nu_{i_k\rightarrow j_k}\hat{B}_{i_{k+1}}.
    \label{simpleCgen}
\end{equation}
Their sum gives,
\begin{align}
    \sum^{+\infty}_{k=0}\hat{C}_{k+1}(\hat{X},\hat{Y})
    &=&\sum^{+\infty}_{k=0}\frac{(-1)^k}{(k+1)!}\sum_{\stackrel{1\leq j_0,i_{k+1}\leq N,\ i_1\rightarrow j_1,\cdots, i_k\rightarrow j_k\in\mathcal{D}_{N,2}}{j_0=i_1, \ldots, j_k=i_{k+1}}}\mu_{j_0}\nu_{i_1\rightarrow j_1}\cdots\nu_{i_k\rightarrow j_k}\hat{B}_{i_{k+1}}.\nonumber\\
    \label{inter2}
\end{align}

By introducing a $N\times N$ symmetric matrix $M$ such that:

\begin{align}
    \forall i\neq j,\ M_{ij}&=&\nu_{i\rightarrow j}\nonumber\\
    M_{ii}&=&0,
\end{align}

as well as $N$-component vectors, $\vec{\mu}\equiv(\mu_{i})_i$, and, $\vec{\hat{B}}\equiv(\hat{B}_{i})_i$,  by noticing that:

\begin{equation}
    (M^{k})_{i_{1}j_{k}}=\sum_{\stackrel{1\leq j_{1},\cdots,i_{k}\leq N}{j_{1}=i_{2},\ldots,j_{k-1}=i_{k}}}\nu_{i_{1}\rightarrow j_{1}}\ldots\nu_{i_{k}\rightarrow j_{k}},
\end{equation}

one can cast Eq.(\ref{inter2}) into:

\begin{equation}
    \sum_{k=0}^{+\infty}\hat{C}_{k+1}(\hat{X},\hat{Y})=\sum_{k=0}^{+\infty}\frac{(-1)^{k}}{(k+1)!}\vec{\mu}\cdot M^{k}\cdot \vec{\hat{B}}.
\end{equation}
Then, if the matrix $M$ is invertible, denoting by  $\mathbb{I}$ the $(N\times N)$-identity matrix,
\begin{equation}
\label{matrixformgeneralcase}
    \sum_{k=0}^{+\infty}\hat{C}_{k+1}(\hat{X},\hat{Y})=\vec{\mu}\cdot\left(\mathbb{I}-\exp(-M)\right) M^{-1}\cdot\vec{\hat{B}}\ .
\end{equation}
It should be stressed out that $M$ is not always invertible, even if all the $\nu_{i\rightarrow j}$ are non-zero and  $N$ as small as $4$. A more general formulation, valid in all cases, is addressed in Appendix B. However, equation (\ref{matrixformgeneralcase}) is easy to handle, and since the set of $M$-matrices with non-zero determinant is a dense set, the optimization of the $\mu_{ij}$ parameters can be restricted to invertible $M$-matrices without loss of accuracy. So, assuming $M$ invertible, a practical expression for the Zassenhaus decomposition is:
\begin{align}
    \exp(\hat{X}+\hat{Y})=\prod\limits_{1\leq i<j\leq N}\exp(\mu_{ij}\hat{A}_{ij})\prod\limits_{1\leq k\leq N}\exp\left(\vec{\mu}\cdot\left(\mathbb{I}-\exp(-M)\right)\cdot M^{-1}_k\hat{B}_k\right),
    \label{general-case-exp}
\end{align}
where $M^{-1}_l$ is the $l^{th}$ column vector of $M^{-1}$.\\

For $N=2$, $M^{-1}=\frac{1}{\mu_{12}^2}M$, and $\exp(-M)=\begin{pmatrix}
\cosh{\mu_{12}}&-\sinh{\mu_{12}}\\-\sinh{\mu_{12}}&\cosh{\mu_{12}}    
\end{pmatrix}$, so that Eq.(\ref{zassenhaus-2pairs-ext}) can be recovered (with no extra $\hat{B}_{k}$).\\

As in the $N=2$ case, it is interesting to consider the limit where the $\mu_{ij}$'s tend to zero, since they are supposed to be small parameters. Then, we have $M\rightarrow 0$, and $\lim\limits_{M\to 0}\vec{\mu}\cdot\left(\mathbb{I}-\exp(-M)\right)\cdot M^{-1}_k=\lim\limits_{M\to 0}\vec{\mu}\cdot M\cdot M^{-1}_k=\mu_k$, so that, $\lim\limits_{M\to 0}\exp\left(\vec{\mu}\cdot\left(\mathbb{I}-\exp(-M)\right)\cdot M^{-1}_k\hat{B}_k\right)=\exp\left(\mu_k\hat{B}_k\right)$. This implies that the Jacobian of the change of variables $\mu_{ij}\rightarrow\mu_{ij},\mu_k\rightarrow\vec{\mu}\cdot\left(\mathbb{I}-\exp(-M)\right)\cdot M^{-1}_k$ is non zero. So, in practice, one can equivalently and more simply optimize the $\gamma_k\equiv\vec{\mu}\cdot\left(\mathbb{I}-\exp(-M)\right)\cdot M^{-1}_k$'s, instead of the $\mu_{k}$'s. If the dominant contributions, in some UCC applications, come from excitation operators satisfying the no-mixed adjoint property, this may also explain why optimizing the parameters after Trotterization actually produces insignificant errors, as observed in Refs.~\cite{Grimsley2019,Sokolov2020}.\\

Each exponential operator in the right-hand side of Eq. (\ref{general-case-exp}) can be implemented by Givens gates, as in the case $N=2$. Their total number will be (at most) $\frac{N(N-1)}{2}$ for the $\hat{A}_{jk}$'s and $N$  for the $\hat{B}_{l}$'s, so $\frac{N(N+1)}{2}$ altogether. 

\section{Conclusion}
We have shown that the Zassenhaus decomposition for the exponential of the sum of two non-commuting operators, simplifies drastically when these operators satisfy a simple condition, called the no-mixed adjoint property.   An important application to a unitary pair Coupled Cluster method has been presented. The simple formula obtained in this work, for the exponential of a cluster operator built from non-commuting mono-excitations and broken-pair di-excitations, shows that, within a change of variables for the optimized parameters, the exponential can be decomposed exactly into a finite product of elementary Givens quantum gates, just as if the excitation operators were commuting.  Alternatively, the formula demonstrates that exact results can be obtained if parameter optimization is performed after $1$-step Trotterization i.e. for disentangled UCC~\cite{Evangelista2019}. Note that, in contrast with Ref.~\cite{Evangelista2019}, our exact product of particle-hole one- and two-body unitary transformations is finite, 
even if we add in front of it, a frozen pair operator acting on a Slater determinant. Note also, that the pair operator, $T_{int}$ of Eq.(\ref{pair-op}), does not need to be frozen: its $\lambda_{pq}$ parameters can be optimized together with the $\mu_{ij}$'s and $\mu_k$'s of $T_{ext}$, provided that extra parametrized Givens gates are added upstream in the ansatz preparation circuit. In any case, the number of Givens gates will not exceed that of free parameters.\\

So, the 2D-Block UpCC method being parsimonious in quantum gates, with no approximation, it seems particularly well-adapted to an implementation on NISQ computers. Based on our previous study~\cite{Cassam_hal-05210807}, we believe that, it can capture the most important contributions to electronic correlations in  molecular systems, including strongly correlated ones. Work is ongoing and we hope to report numerical results, very soon.\\

We also hope that the present study can trigger the construction of other ans\"atze, from similar particule-hole operators satisfying the no-mixed adjoint property, in quantum chemistry or in other quantum physical contexts, such as condensed matter, for models based on electron pairs or in nuclear physics for pairs of nucleons.\\
 
\paragraph{Acknowledgements}
We acknowledge the use of IBM Quantum services for this work. The views expressed are those of the authors, and do not reflect the official policy or position of IBM or the IBM Quantum team.
We are grateful to Dr. F. Patras for fruitful discussions and insightful hints.\\

\section*{Statements and Declarations}

\begin{itemize}
\item \textbf{Funding}

This work was supported by the French government through a QuanTEdu-France doctoral contract, within the France 2030 investment plan managed by the National Research Agency (ANR). \\

\item \textbf{Conflict of interest/Competing interests}

The authors have no conflicts of interest to disclose.\\

\item \textbf{Data Availability}

No data are required to support the findings of this study\\

%
%
%
%
%
\end{itemize}


\newpage

\appendix

\section*{Appendices}

\section{Proof of theorem \ref{theorem} }

\subsection{Lemma 1}

Let us first consider the following lemma:

\begin{theorem}[Lemma 1]For a couple of operators $\hat{X}$ and $\hat{Y}$, $p\in\mathbb{N}^{*}$, $q,m\in\mathbb{N}$, $p\ne q$ and $0\leq m\leq p-1$, we have:

\begin{equation}
\label{lemme1}
    [ad^{p}_{\hat{X}}\hat{Y},ad^{q}_{\hat{X}}\hat{Y}]=\sum^{m+1}_{n=0}(-1)^{m+1-n}\binom{m+1}{n}ad^{n}_{\hat{X}}[ad^{p-m-1}_{\hat{X}}\hat{Y},ad^{q+m+1-n}_{\hat{X}}\hat{Y}]
\end{equation}
\end{theorem}

\textbf{\underline{Proof:}}\\

We prove the lemma by induction on $m$. For $m=0$, Eq.(\ref{lemme1}) gives:

\begin{equation}
\label{m0lemme1}
    [ad^{p}_{\hat{X}}\hat{Y},ad^{q}_{\hat{X}}\hat{Y}]=ad_{\hat{X}}[ad^{p-1}_{\hat{X}}\hat{Y},ad^{q}_{\hat{X}}\hat{Y}]-[ad^{p-1}_{\hat{X}}\hat{Y},ad^{q+1}_{\hat{X}}\hat{Y}]
\end{equation}

Some simple algebra shows that this equality is verified:

\begin{align*}
    [ad^{p}_{\hat{X}}\hat{Y},ad^{q}_{\hat{X}}\hat{Y}]&=[[\hat{X},ad^{p-1}_{\hat{X}}\hat{Y}],ad^{q}_{\hat{X}}\hat{Y}]\\
    &=[\hat{X}(ad^{p-1}_{\hat{X}}\hat{Y}),ad^{q}_{\hat{X}}\hat{Y}]-[(ad^{p-1}_{\hat{X}}\hat{Y})\hat{X},ad^{q}_{\hat{X}}\hat{Y}]\\
    &=\hat{X}[ad^{p-1}_{\hat{X}}\hat{Y},ad^{q}_{\hat{X}}\hat{Y}]+(ad^{q+1}_{\hat{X}}\hat{Y})(ad^{p-1}_{\hat{X}}\hat{Y})-(ad^{p-1}_{\hat{X}}\hat{Y})(ad^{q+1}_{\hat{X}}\hat{Y})-[ad^{p-1}_{\hat{X}}\hat{Y},ad^{q}_{\hat{X}}\hat{Y}]\hat{X}\\
    &=[\hat{X},[ad^{p-1}_{\hat{X}}\hat{Y},ad^{q}_{\hat{X}}\hat{Y}]]-[ad^{p-1}_{\hat{X}}\hat{Y},ad^{q+1}_{\hat{X}}\hat{Y}]\\
    &=ad_{\hat{X}}[ad^{p-1}_{\hat{X}}\hat{Y},ad^{q}_{\hat{X}}\hat{Y}]-[ad^{p-1}_{\hat{X}}\hat{Y},ad^{q+1}_{\hat{X}}\hat{Y}].
\end{align*}

Suppose now that Eq.(\ref{lemme1}) is satisfied for $m<p-1$. Injecting Eq.(\ref{m0lemme1}) in Eq.(\ref{lemme1}), we obtain:

\begin{align*}
    [ad^{p}_{\hat{X}}\hat{Y},ad^{q}_{\hat{X}}\hat{Y}]&=\sum^{m+1}_{n=0}(-1)^{m+1-n}\binom{m+1}{n}ad^{n}_{\hat{X}}(ad_{\hat{X}}[ad^{p-m-2}_{\hat{X}}\hat{Y},ad^{q+m+1-n}_{\hat{X}}\hat{Y}]\\&-[ad^{p-m-2}_{\hat{X}}\hat{Y},ad^{q+m+1-n+1}_{\hat{X}}\hat{Y}])\\&=\sum^{m+1}_{n=0}(-1)^{m+1-n}\binom{m+1}{n}ad^{n+1}_{\hat{X}}[ad^{p-m-2}_{\hat{X}}\hat{Y},ad^{q+m+1-n}_{\hat{X}}\hat{Y}]\\&-\sum^{m+1}_{n=0}(-1)^{m+1-n}\binom{m+1}{n}ad^{n}_{\hat{X}}[ad^{p-m-2}_{\hat{X}}\hat{Y},ad^{q+m+2-n}_{\hat{X}}\hat{Y}].
\end{align*}

Setting $u=n+1$ in the first term, then renaming $u$ as $n$, we get:

\begin{align*}
    [ad^{p}_{\hat{X}}\hat{Y},ad^{q}_{\hat{X}}\hat{Y}]&=\sum^{m+2}_{n=1}(-1)^{m+2-n}\binom{m+1}{n-1}ad^{n}_{\hat{X}}[ad^{p-m-2}_{\hat{X}}\hat{Y},ad^{q+m+2-n}_{\hat{X}}\hat{Y}]-\\&\sum^{m+1}_{n=0}(-1)^{m+1-n}\binom{m+1}{n}ad^{n}_{\hat{X}}[ad^{p-m-2}_{\hat{X}}\hat{Y},ad^{q+m+2-n}_{\hat{X}}\hat{Y}]\\&=\sum^{m+1}_{n=1}(-1)^{m+2-n}\left[\binom{m+1}{n-1}+\binom{m+1}{n}\right ]ad^{n}_{\hat{X}}[ad^{p-m-2}_{\hat{X}}\hat{Y},ad^{q+m+2-n}_{\hat{X}}\hat{Y}]\\&+ad^{m+2}_{\hat{X}}[ad^{p-m-2}_{\hat{X}}\hat{Y},ad^{q}_{\hat{X}}\hat{Y}]-(-1)^{m+1}[ad^{p-m-2}_{\hat{X}}\hat{Y},ad^{q+m+2}_{\hat{X}}\hat{Y}].
\end{align*}

This can be re-condensed with the help of Pascal's triangle,

\begin{align*}
    [ad^{p}_{\hat{X}}\hat{Y},ad^{q}_{\hat{X}}\hat{Y}]&=\sum^{m+1}_{n=1}(-1)^{m+1-n+1}\binom{m+2}{n}ad^{n}_{\hat{X}}[ad^{p-m-2}_{\hat{X}}\hat{Y},ad^{q+m+1-n+1}_{\hat{X}}\hat{Y}]\\&+ad^{m+2}_{\hat{X}}[ad^{p-m-2}_{\hat{X}}\hat{Y},ad^{q}_{\hat{X}}\hat{Y}]+(-1)^{m+2}[ad^{p-m-2}_{\hat{X}}\hat{Y},ad^{q+m+2}_{\hat{X}}\hat{Y}]\\&=\sum^{m+2}_{n=0}(-1)^{m+2-n}\binom{m+2}{n}ad^{n}_{\hat{X}}[ad^{p-m-2}_{\hat{X}}\hat{Y},ad^{q+m+2-n}_{\hat{X}}\hat{Y}]
\end{align*}

Eq.(\ref{lemme1}) is thus verified for $m+1$. Hence, the lemma, which leads to the following corollary.\\

\subsection{Corollary}

\begin{theorem}[Corollary]Let $(\hat{X},\hat{Y})$ satisfying the no-mixed adjoint property,
 $\forall p,q\in\mathbb{N}$. Then,

\begin{equation}
    [ad^{p}_{\hat{X}}\hat{Y},ad^{q}_{\hat{X}}\hat{Y}]=0.
\label{conseqlemme1}
\end{equation}
\end{theorem}

\textbf{\underline{Proof:}}\\
If p=q, the result is obvious. So, we can choose $p>q$ without loss of generality,
and apply lemma 1 with $m=p-1$:
\begin{equation}
    [ad^{p}_{\hat{X}}\hat{Y},ad^{q}_{\hat{X}}\hat{Y}]=\sum^{p}_{n=0}(-1)^{p-n}\binom{p}{n}ad^{n}_{\hat{X}}ad_{\hat{Y}}ad^{q+p-n}_{\hat{X}}\hat{Y}=0
\end{equation}
because of the no-mixed adjoint property.\\

We are now able to prove our main theorem.\\

\subsection{Proof of the main theorem:}

We recall the recursive relation established in~\cite{Casas2012} to calculate the successive $\hat{C}_{n}(\hat{X},\hat{Y})$. For $n=1,2,3$, $\hat{C}_{n}$ is given by:

\begin{equation}
\label{crecurrenceinitial}
    \hat{C}_{n+1}=\frac{1}{n+1}\hat{f}_{1,n}\ ,
\end{equation}

with, for $n\in\mathbb{N}^{*}$,

\begin{equation}
    \hat{f}_{1,n}=\sum^{n}_{j=1}\frac{(-1)^{n}}{j!(n-j)!}ad^{n-j}_{\hat{Y}}ad^{j}_{\hat{X}}\hat{Y}\ .
\label{finitial}
\end{equation}

For $n\geqslant4$, another formula is required:

\begin{equation}
\label{crecurrence}
    \hat{C}_{n+1}=\frac{1}{n+1}\hat{f}_{[\frac{n}{2}],n}
\end{equation}

where $[.]$ denotes the integer part and $\hat{f}_{m,n}$, for $n\geqslant m\geqslant2$, is given by: 

\begin{equation}
\label{frecurrence}
    \hat{f}_{m,n}=\sum^{[\frac{n}{m}]-1}_{j=0}\frac{(-1)^{j}}{j!}ad^{j}_{\hat{C}_{m}}\hat{f}_{m-1,n-mj}
\end{equation}

Let us show  by induction on $m$ that  $\forall n\geqslant m\geqslant1$, $\hat{f}_{m,n}=\frac{(-1)^{n}}{n!}ad^{n}_{\hat{X}}\hat{Y}$ (independently of $m$) when the no-mixed adjoint property is satisfied.\\

For $m=1$, the result is obtained immediately by using the no-mixed adjoint property in Eq.(\ref{finitial}), the only non-zero term corresponding to $j=n$.


Suppose that the result is true up to $m\in\mathbb{N}^{*}$. We have $\forall n\geqslant m+1$ by Eq.(\ref{frecurrence}):

\begin{align}
    \hat{f}_{m+1,n}&=&\sum^{[\frac{n}{m+1}]-1}_{j=0}\frac{(-1)^{j}}{j!}ad^{j}_{\hat{C}_{m+1}}\hat{f}_{m,n-(m+1)j}\nonumber\\
&=&\hat{f}_{m,n}+\sum^{[\frac{n}{m+1}]-1}_{j=1}\frac{(-1)^{j}}{j!}ad^{j-1}_{\hat{C}_{m+1}}ad^{1}_{\hat{C}_{m+1}}\hat{f}_{m,n-(m+1)j}.
\label{inter1}
\end{align}

Considering the right-hand side factors appearing in the summation and substituting Eq.(\ref{crecurrence}), we have:

\begin{align*}
    ad^1_{\hat{C}_{m+1}}\hat{f}_{m,n-(m+1)j}=[\hat{C}_{m+1},\hat{f}_{m,n-(m+1)j}]&=\frac{1}{m+1}[\hat{f}_{[\frac{m}{2}],m},\hat{f}_{m,n-(m+1)j}]\\&\propto[ad^{m}_{\hat{X}}\hat{Y},ad^{n-(m+1)j}_{\hat{X}}\hat{Y}],
\end{align*}
where we have used the induction hypothesis in the last line.
From the corollary Eq.(\ref{conseqlemme1}), we deduce $ad^1_{\hat{C}_{m+1}}\hat{f}_{m,n-(m+1)j}=0$, so that Eq.(\ref{inter1}) reduces to:

\begin{equation*}
    \hat{f}_{m+1,n}=\hat{f}_{m,n}=\frac{(-1)^{n}}{n!}ad^{n}_{\hat{X}}\hat{Y}
\end{equation*}

which is the expected result.\\

The first three terms are according to Eq.(\ref{crecurrenceinitial}):

\begin{align*}
    \hat{C}_{2}(\hat{X},\hat{Y})&=-\frac{1}{2}[\hat{X},\hat{Y}]\\
    \hat{C}_{3}(\hat{X},\hat{Y})&=\frac{1}{6}[\hat{X},[\hat{X},\hat{Y}]]\\
    \hat{C}_{4}(\hat{X},\hat{Y})&=-\frac{1}{24}[\hat{X},[\hat{X},[\hat{X},\hat{Y}]]].
\end{align*}
Then, for $n\geq 5$, since $\hat{f}_{[\frac{n-1}{2}],n-1}=\hat{f}_{1,n-1}=\frac{(-1)^{n-1}}{(n-1)!}ad^{n-1}_{\hat{X}}\hat{Y}$, Eq.(\ref{crecurrence}) shows that, actually, the following formula holds true for all $n\geqslant2$.
\begin{equation}
    \hat{C}_{n}(\hat{X},\hat{Y})=\frac{(-1)^{n-1}}{n!}ad^{n-1}_{\hat{X}}\hat{Y}.
\end{equation}

\newpage

\section{Simplified Zassenhaus formula in terms of  $*$-product on a free algebraic structure}

To reach a simplified expression of Zassenhaus formula similar to the $(N=2)$-case, for operators satisfying the no-mixed adjoint property, we define a $*$-product on the free algebra generated by the $\nu_{i\rightarrow j}$'s:
\begin{equation}
    \forall i,j,k,l \quad \nu_{i\rightarrow j}*\nu_{k\rightarrow l}=\delta_{jk}\ \nu_{i\rightarrow j}\nu_{j\rightarrow l},
\end{equation}
where $\delta_{jk}$ is a Kronecker-like delta giving the empty symbol instead of the concatenated one, $\nu_{i\rightarrow j}\nu_{j\rightarrow l}$, when $j\neq k$.
This operation is non-commutative. However, it is associative and distributive over $+$.
Considering the indices of single index $\mu$ coefficients and $\hat{B}$-operators as vertices, we extend the definition of the $*$-product to a wider free algebraic structure, where every word starts with a $\mu$ symbol and ends with a $\hat{B}$ symbol, (it can be seen as the tensor product of the $\mu$-symbol free module with the $\nu$-symbol free algebra, tensor product with the $\hat{B}$ symbol free module):
\begin{align}
    \forall i,j,k,l \quad \mu_{i}*\nu_{k\rightarrow l}&=&\delta_{ik}\ \mu_{i}\nu_{i\rightarrow l}\\
    \nu_{i\rightarrow j}*\hat{B}_{k}&=&\delta_{jk}\ \nu_{i\rightarrow j}\hat{B}_{j}\\
    \mu_{i}*\hat{B}_{k}  &=&\delta_{ik}\ \mu_{i}\hat{B}_{i}
\end{align}
Then, denoting $\pi$ the morphism projecting the free algebra on the real numbers, Eq.(\ref{inter2}) can be rewritten as 
\begin{align}
    \sum^{+\infty}_{k=0}\hat{C}_{k+1}(\hat{X},\hat{Y})&=&\sum^{+\infty}_{k=0}\frac{(-1)^k}{(k+1)!}\sum_{1\leq j_0,i_{k+1}\leq N,\ i_1\rightarrow j_1,\cdots, i_k\rightarrow j_k\in\mathcal{D}_{N,2}}\pi(\mu_{j_0}*\nu_{i_1\rightarrow j_1}*\cdots *\nu_{i_k\rightarrow j_k}*\hat{B}_{i_{k+1}})\nonumber\\
    &=&\sum_{1\leq i,l\leq N}\pi\left(\mu_{i}*\sum^{+\infty}_{n=0}\frac{(-1)^n}{(n+1)!}\left(\sum_{j\rightarrow k\in\mathcal{D}_{N,2}}\nu_{j\rightarrow k}\right)^{*n}*\hat{B}_{l}\right),
    \label{inter3}
\end{align}
where we have used the notation $x^{*n}\equiv\underbrace{x*\cdots *x}_n$. We set in the same spirit:
\begin{align}
  ^{*}e^x&=&\sum^{+\infty}_{n=0}\frac{x^{*n}}{n!} ,
\end{align}
defining a $*$-exponential function.
To make this function appear in Eq.(\ref{inter3}), we need to define an inverse operation to the $*$-product. We define the $^{*-1}$-operation by, $\forall i,j,k,l\in\{1,\ldots , N\}$:
\begin{align}
\mathrm{for}\ \nu_{i\rightarrow j} &=&0 , \qquad  \nu_{i\rightarrow j}( \nu_{k\rightarrow l})^{*-1}=0,\qquad \forall \nu_{k\rightarrow l}\\
\mathrm{for}\ \nu_{i\rightarrow j} &\neq&0 , \qquad  \nu_{i\rightarrow j}( \nu_{k\rightarrow l})^{*-1}=\frac{\delta_{il}\delta_{jk}}{\nu_{i\rightarrow j}}.
\end{align}
Then, omitting the $\pi$ projection to alleviate the notation, (that is identifying implicitly a word with its projection on the field of real numbers), 
\begin{align}
    &\sum\limits^{+\infty}_{k=0}&\hat{C}_{k+1}(\hat{X},\hat{Y})=\sum_{1\leq i,l\leq N}\mu_{i}*\sum^{+\infty}_{n=0}\frac{(-1)^n}{(n+1)!}\left(\sum_{j\rightarrow k\in\mathcal{D}_{N,2}}\nu_{j\rightarrow k}\right)^{*(n+1)}*\left(\sum_{j\rightarrow k\in\mathcal{D}_{N,2}}\nu_{j\rightarrow k}^{*-1}\right)*\hat{B}_{l}\nonumber\\
    &=&\sum_{1\leq i,l\leq N}\mu_{i}*\left(1-^*e^{-\sum\limits_{j\rightarrow k\in\mathcal{D}_{N,2}}\nu_{j\rightarrow k}}\right)*\left(\sum_{j\rightarrow k\in\mathcal{D}_{N,2}}\nu_{j\rightarrow k}^{*-1}\right)*\hat{B}_{l}\nonumber\\
    \label{cas-gen}
\end{align}
and the Zassenhaus decomposition is:
\begin{align}
   & \exp(\hat{X}+\hat{Y})=&\prod\limits_{1\leq i<j\leq N}\exp(\mu_{ij}\hat{A}_{ij})\prod\limits_{1\leq k\leq N}\exp\left(\sum\limits_{1\leq l\leq N}\mu_{l}*\left(1-^*e^{-\sum\limits_{p\rightarrow q\in\mathcal{D}_{N,2}}\nu_{p\rightarrow q}}\right)\right.\nonumber\\&&\left.*\left(\sum\limits_{p\rightarrow q\in\mathcal{D}_{N,2}}\nu_{p\rightarrow q}^{*-1}\right)*\hat{B}_{k}\right).
\end{align}
This shows that, even in the general case one needs to implement only $\frac{N(N-1)}{2}$ quantum circuits for the $\hat{A}_{jk}$'s and $N$ quantum circuits for the $\hat{B}_{l}$'s, so at most $\frac{N(N+1)}{2}$ altogether.

\newpage

\section*{Figures}

\begin{figure}[htpb]
    \centering

\begin{subfigure}{1.0\textwidth}
\includegraphics[width=\textwidth]{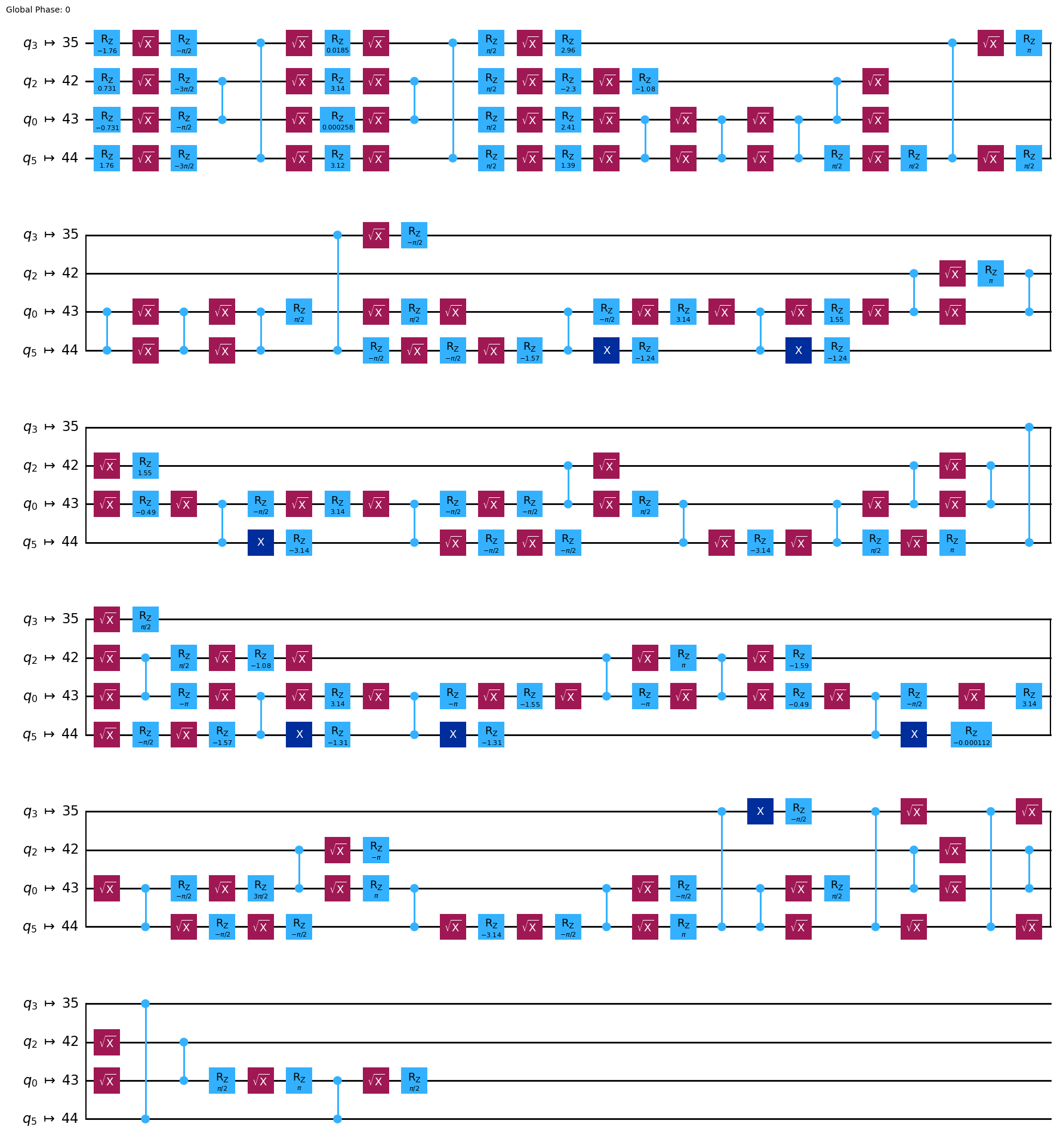}
    \caption{4-qubit Givens gate circuit}
    \label{circ1-trans}
\end{subfigure}
\end{figure}
\begin{figure}[htpb]
\ContinuedFloat
\begin{subfigure}{1.0\textwidth}
\includegraphics[width=\textwidth]{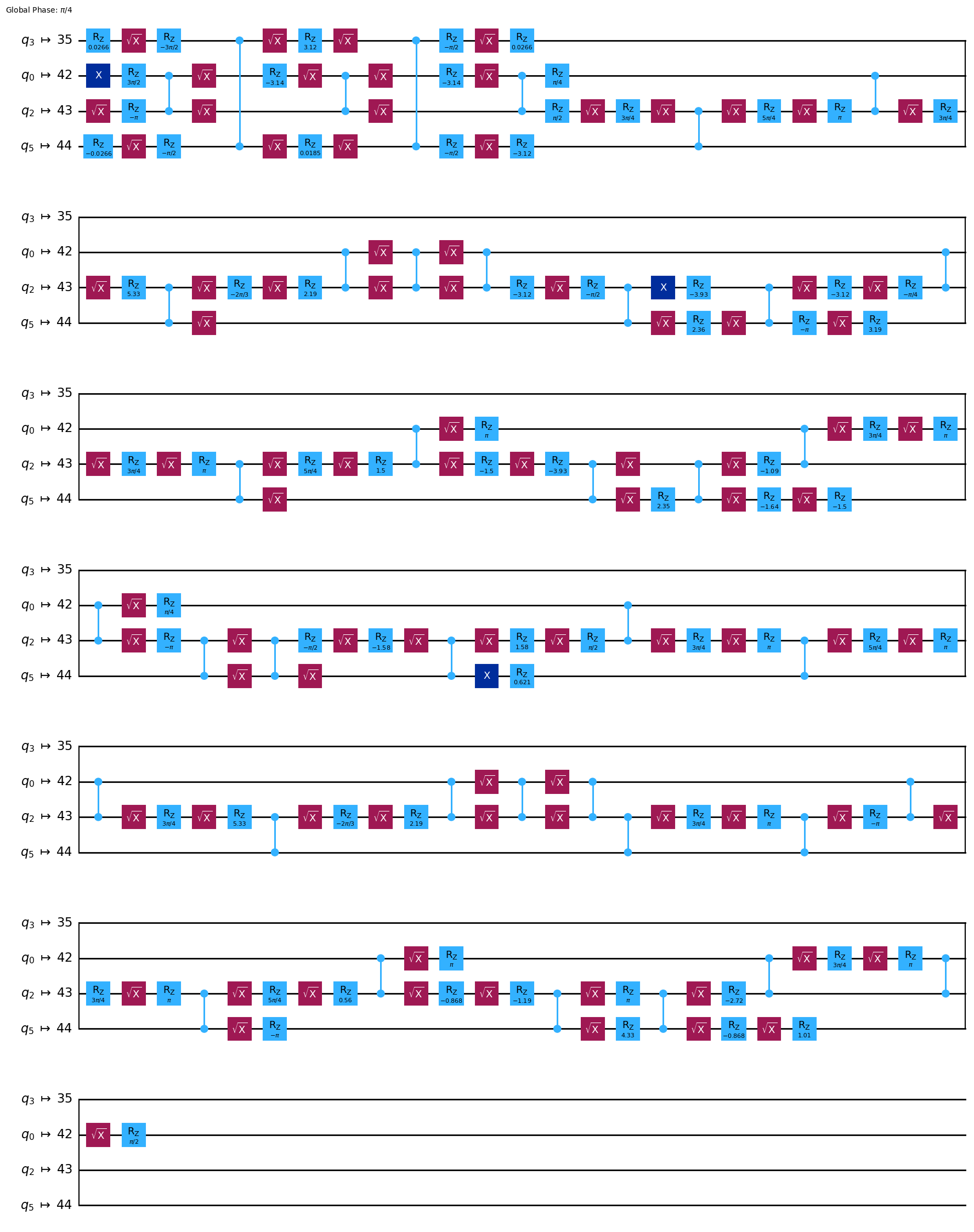}
    \caption{Conditional 2-qubit Givens gate circuit}
    \label{circ2-trans}
\end{subfigure}

\caption{Comparison of the  transpiled circuits of Fig.1 on Heron Torino IBM quantum computer. Transpilations have been performed at level 3 of optimization for both circuits. The upper circuit corresponds to the four-qubit Givens gate and has a shorter depth than the lower one, corresponding to the conditional two-qubit Givens gate. However, the total count of two-qubit CZ basic gates is larger in the former than in the latter ($44$ against $39$). The circuit drawings are exported from the IBM Quantum Qiskit packages.}
\label{real-circuits}
\end{figure}

\end{document}